\documentclass{llncs}

\usepackage{amsmath}
\usepackage{graphicx}
\usepackage{url}
\usepackage{hyperref}
\usepackage{cite}

\newcommand{\types}{\ensuremath{\mathcal{T}}}
\newcommand{\edges}{\ensuremath{\mathcal{E}}}
\newcommand{\attr}{\ensuremath{\mathcal{A}}}
\newcommand{\vals}{\ensuremath{\mathcal{V}}}

\newcommand{\eg}{\emph{e.g.}}

\newcommand{\ie}{\emph{i.e.}}
\newcommand{\etc}{\emph{etc}}
\newcommand{\etal}{\emph{et al}}

\newcommand{\scb}{\ensuremath{\mathsf{SC}}}
\newcommand{\lcb}{\ensuremath{\mathsf{LC}}}
\newcommand{\unk}{\ensuremath{\mathsf{U}}}
\newcommand{\ltb}{\ensuremath{\mathsf{LT}}}
\newcommand{\stb}{\ensuremath{\mathsf{ST}}}

\title{Representing Network Trust and Using It to Improve Anonymous Communication\thanks{Supported in part by NSF grant 1016875 and in part by the Defense Advanced Research Agency (DARPA) and SPAWAR Systems Center Pacific, Contract No.\ N66001-11-C-4018.  Work at NRL also supported by ONR.}}

\author{Aaron D.\ Jaggard\inst{1} \and Aaron Johnson\inst{1} \and Paul Syverson\inst{1} \and Joan Feigenbaum\inst{2}}

\institute{U.S.\ Naval Research Laboratory \email{$\{$aaron.jaggard,aaron.m.johnson,paul.syverson$\}$@nrl.navy.mil}
\and Yale University \email{joan.feigenbaum@yale.edu}}

\begin{document}

\maketitle

\begin{abstract}
Motivated by the effectiveness of correlation attacks against Tor, the censorship arms race, and
observations of malicious relays in Tor, we propose that Tor users capture their trust in 
network elements using 
probability distributions over the sets of elements observed by network
adversaries. We present a modular system that allows users to efficiently 
and conveniently create such distributions and use them to improve their security. The major components of this
system are (\emph{i}) an ontology of network-element types that represents the main
threats to and vulnerabilities of anonymous communication over Tor, (\emph{ii})
a formal language that allows users to naturally express trust beliefs about network elements, and (\emph{iii}) a conversion procedure that takes the ontology, public information about the network, and user beliefs written in the
trust language and produce a Bayesian Belief Network that represents the
probability distribution in a way that is concise and easily sampleable.  We also present preliminary experimental results that show the distribution produced by our system can improve security when employed by users; further improvement is seen when the system is employed by both users and services.
\end{abstract}

\textbf{Keywords:} Tor, Trust, Bayesian Belief Network

\section{Introduction}

Tor and its users currently face serious security risks from adversaries positioned to observe traffic into and out of the Tor network.  Large-scale deanonymization
has recently been shown feasible~\cite{usersrouted-ccs13} for a patient adversary that
controls some network infrastructure or Tor relays. Such adversaries are a real and growing threat,
as demonstrated by the ongoing censorship arms race~\cite{eg12cordon} and recent observations of
malicious Tor relays~\cite{wl14arxiv}.  In light of these and other threats, we propose an approach to representing and using trust in order to improve anonymous communication in Tor.  Trust information can be used to inform path selection by Tor users and the location of services that will be accessed through Tor, in both cases strengthening the protection provided by Tor.  A better understanding of trust-related issues will also inform the future evolution of Tor, both the protocol itself and its network infrastructure.

Attacks on Tor users and services include first--last correlation~\cite{strl01correlation}, in which an adversary correlates traffic patterns between the client and a guard with traffic patterns between a Tor exit and a network destination in order to link the client to her destination.  They also include more recently identified attacks on a single end of a path such as fingerprinting users~\cite{czjj12ccs} or
services~\cite{Biryukov-2013}. With trust information, users could choose trusted paths through the Tor network and services could choose server locations with trusted paths into the network in order
to reduce the chance of these attacks.  Other work~\cite{js09csf,jsdm11ccs} has considered the use of trust to improve security in Tor.  The work presented here is novel in that (\emph{i}) it considers trust in network elements generally and not just Tor relays and (\emph{ii}) it considers more general adversary distributions than in previous work.

The system we describe here is designed to produce a distribution on the sets of network locations that might be compromised by a single adversary.  In the case of multiple, non-colluding adversaries, multiple distributions could be produced.  These distribution can then be used, \eg, as part of a user's path-selection algorithm in Tor.  In constructing our preliminary experiments, we suggest how our distributions may be used in this way.  Here, we capture these distributions using Bayesian Belief Networks (BBNs; see, \eg, \cite{halpern03book}).

The contribution of this work is the proposal of a modular system that (\emph{i})~allows users to express beliefs about the structure and trustworthiness of the network, (\emph{ii})~uses information about the network, modified according to the user-provided structural information, to produce a ``world'' that captures how compromise is propagated through the network, and (\emph{iii})~combines this world with the user's trust beliefs to produce a BBN representing a distribution on the sets of network elements that an adversary might compromise.  As part of our contribution, we present results of proof-of-concept experiments.  These show that users can employ our system to reduce their risk of first--last correlation; this risk is reduced even further when our system also informs the locations that services choose for their servers.

The body of this paper provides a high-level view of our system, starting with an overview of its 
operation.  In addition to describing what the system provides and how it is combined with user 
beliefs to produce a BBN, we discuss some issues related to users' trust beliefs.  
We then present our experimental results and sketch ongoing and future work.  
As noted throughout, additional details and examples are provided in the appendices.

\section{System Overview}

We survey our system, which is largely modular.  This allows it to be extended as new types of trust 
information are identified as important, \etc.  The system comes with an ontology that describes 
types of network elements (\eg, Autonomous System (AS) and relay-operator types), the relationships 
between them that capture correlated compromise by an adversary, and attributes of these things.  
Using the ontology and various published data about the network, the system creates a preliminary 
``world'' populated by real-world instances of the ontology types (\eg, specific ASes and relay 
operators).  The world also includes relationship instances that reflect which particular type 
instances are related in ways suggested by the ontology.  User-provided information may include 
revisions to this system-generated world, including the addition of types not included in the 
provided ontology and instances of both ontology-provided and user-added types.  The user may also 
enrich the information about the effects of compromise (adding, \eg, budget constraints or some 
correlations).  The user also provides beliefs about her trust in particular network elements and 
how her trust in network elements is affected by different attributes of those elements.  This 
user-provided information is used, together with the edited world, to create a Bayesian Belief 
Network (BBN) that encodes the probability distribution on the adversary's location arising from
the user's trust beliefs.  The BBN can, for example, provide samples from the distribution of the
Tor relays and Tor ``virtual links'' (transport-layer connections with Tor relays) that are
observed by the adversary. Appendix~\ref{ap:overview} provides an expanded survey of
the system's architecture.

\section{Ontology and World}

\begin{figure}[htp]
\begin{center}
\includegraphics[width=.9\textwidth]{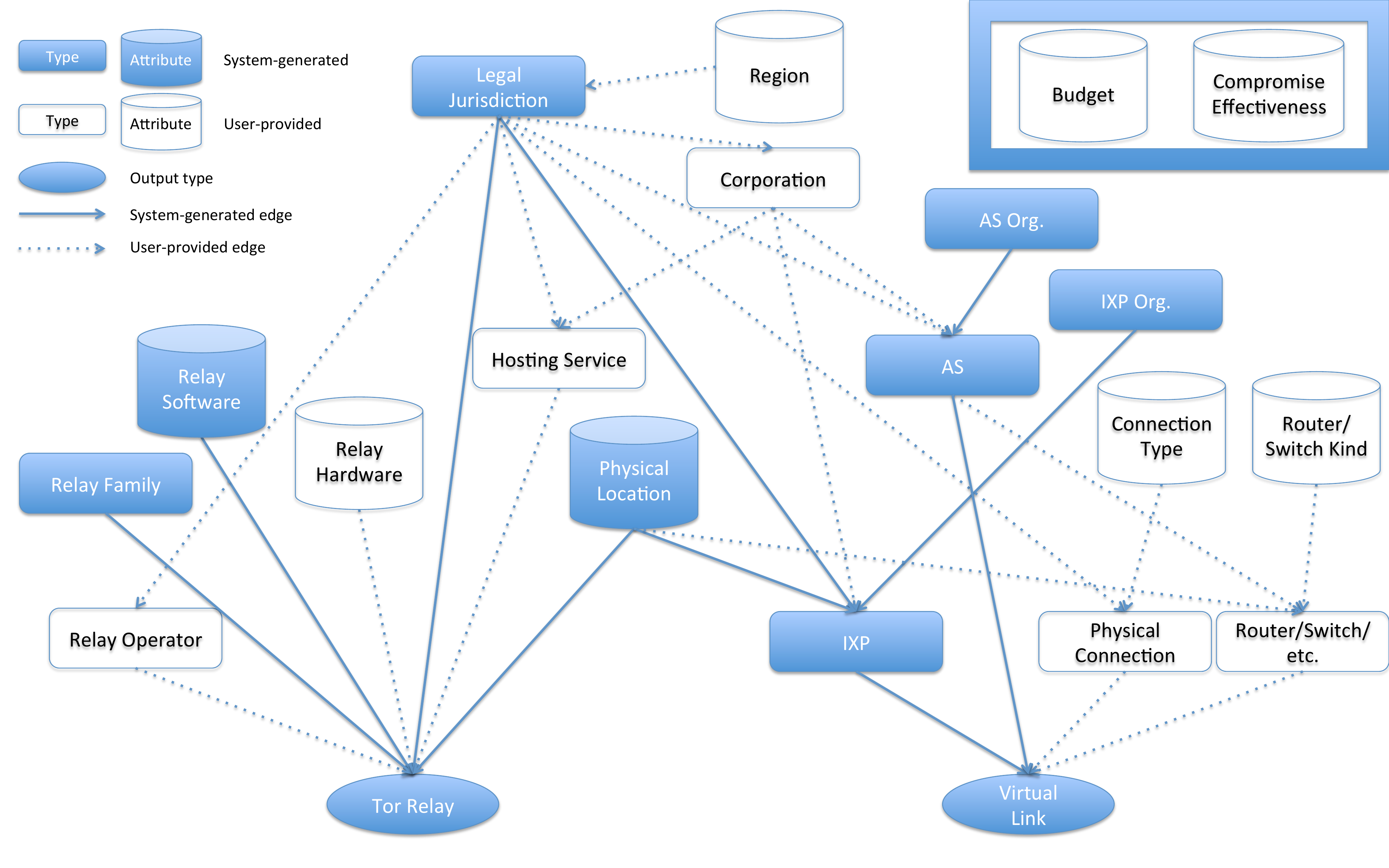}
\end{center}
\caption{Graphical depiction of the system's ontology}\label{fig:ontology}
\end{figure}

Figure~\ref{fig:ontology} shows the elements of our ontology.  Rounded rectangles are types, and ovals are output types.  Cylinders are attributes; with the exception of Relay Software and Physical Location, which the user may modify, these are provided by the user.  The user may also provide new attributes.  Directed edges show expected relationships between types.  For example, the edge from the ``AS'' type to the ``Router/switch'' type indicates that we expect that the compromise of an AS will likely contribute to the compromise of one or more routers and switches.

Other ontologies may modularly replace the one described here if they satisfy the assumptions described in App.~\ref{ap:ontology}.  That appendix also provides details about the elements of the ontology we use here.

The system constructs a preliminary world including instances of Tor relays, relay families,
ASes, Internet Exchange Points (IXPs), AS and IXP organizations, virtual links between every AS and Tor relay, and countries (as legal jurisdictions).  The system assigns to relevant instances the relay-software (from Tor descriptors~\cite{tor-pathspec}) and physical-location (from, \eg, the MaxMind GeoIP database~\cite{maxmind-geoip}) attributes. The
system also creates relationships from families to their relays, countries to the relays and
IXPs they contain, AS and IXP organizations to their members, and ASes and IXPs to the virtual links on which they appear (determined by an AS-level routing map~\cite{usersrouted-ccs13}). 
A fuller description of this part of the process is given in App.~\ref{app:sysworld}.

\section{Beliefs and BBNs}

The user may provide various data to inform the operation of the system.  However, many users may not wish to do this, and the system includes a default belief set designed to provide good security for average users.  In Sect.~\ref{sec:exp} we describe a possible default belief set motivated by the discussion of trust in Sect.~\ref{sec:trust}.  For simplicity, we refer to beliefs as being provided by the user, but wherever they are not, the defaults are used instead.

The user provides structural information that is used to revise the system-generated world.  This may include new types (\eg, law-enforcement treaties) and the addition or removal of type instances and relationships between them (\eg, adding relay operators known to the user). The user may also
define new attributes, change the system-provided attributes, or provide values for empty attributes
(\eg, labeling countries by their larger geographic region).

The user's beliefs may incorporate boolean predicates that are evaluated on instances in the revised world. For example, the user may have increased trust in ASes above a certain size.  We sketch a suitable language for this in App.~\ref{app:predlang}, but this can be replaced with another if desired.

Finally, the user provides beliefs of four types that are used in constructing the BBN from the revised world.  The first two concern the propagation of compromise.  Budget beliefs allow the user to say that an instance $I$ in the edited world has the resources (monetary or otherwise) to compromise $k$ of its children that satisfy some predicate $\mathsf{P}$.  Enforcing this as a hard bound appears to be computationally harder than we are willing to use in the BBN, so we do this in expectation.  Compromise-effectiveness (CE) beliefs allow the user to express some correlations between the compromises of nodes by saying that, if an instance $I$ is compromised, then, with probability $p$, all of $I$'s children satisfying a predicate $\mathsf{P}$ are compromised.  For example, this captures the possibility that a compromised AS compromises all of its routers except those of a particular model, for which the AS has made an error in their (common) configuration file.

The other two belief types concern the likelihood of compromise.  Relative beliefs allow the user to say that instances satisfying a given predicate (\eg, relays running a buggy OS, network links that traverse a submarine cable, or ASes that are small as determined by their number of routers) have a certain probability of compromise.  (In particular, it specifies the probability that they remain uncompromised if they are otherwise uncompromised.)  Absolute beliefs allow the user to say that instances satisfying a given predicate (\eg, the node is an AS and the AS number is 7007) are compromised with a certain probability, regardless of other factors.

The BBN construction from the edited world is described in detail in App.~\ref{app:bbntrans}.  In brief, the nodes from the edited world are copied to the BBN.  Compromise-effectiveness beliefs add nodes to guarantee the correlations (these new nodes are compromised with some probability; if one is compromised, then all of its children are compromised with probability 1).  Other than accounting for these new nodes, the directed edges in the BBN are those from the edited world.  Budget beliefs may further change the probability that compromise is propagated along a directed edge.  The values associated with the relative beliefs that apply to a node are associated with that node.  Unless there is an absolute belief that applies to the node (and would determine the node's compromise probability), the node's probability of compromise is $1-\left(\prod_{p\in S} (1-p)\right)\left(\prod_{q\in R}(1-q)\right)$, where $S$ is the (multi)set of compromise-propagation values associated with the edges from the node's compromised parents and $R$ is the (multi)set of values from the relative beliefs that apply to the node.

\section{Trust}\label{sec:trust}

We now discuss where trust judgments come from by sketching, as a
simple example, the trust rationale behind a Tor trust policy that might
be distributed with client software as a default. Such a policy would be
designed not to offer the best protection to particular classes of
users but to adequately protect most Tor users regardless of where
they are connecting to the network or what their destinations and
behaviors are.

The most useful information about Tor relays for setting a default
level of trust is probably relay longevity. Running a relay in order
to observe traffic at some future time or for persistent observation
of all traffic requires a significant investment of money and possibly
official authorization approval. This is all the more true if the
relay contributes significant persistent capacity to the
network. Further, operators of such relays are typically more
experienced in many senses and thus somewhat less open to external
compromise via hacking. The amount of relay trust is thus usefully
tied to the length of presence in the network consensus, uptime, and
bandwidth.  This approach does not resist a large-budget,
nation-state-scale adversary with authority to monitor relays
persistently, but it will help limit attacks to adversaries with such
persistent capabilities. 

There is no general reason to trust one AS, IXP, \etc., more than
another, but one should not presume that they are all completely
safe. It thus is reasonable to assume the same moderate risk of compromise for
all elements forming the links to the Tor network and between the
relays of the network when creating a default trust policy.

An example of an important non-default case is connecting users to
sensitive destinations that they especially do not want linked to
their location or possibly to their other Tor behaviors. For example,
some users need to connect to sensitive employer hosts, and dissident
bloggers could be physically at risk if seen posting to controversial
sites. These users may have rich trust beliefs (either of their own or
supplied by their organizations) about particular relays, ASes, \etc., based on who runs the relay, hardware, location, \etc.

Note that the average client using a default trust policy may be subject to
errors because the average client will rarely be exactly at the client
average, and all clients may be subject to errors in judgments underlying
a trust policy.

\section{Experimental Results}\label{sec:exp}
As a proof of concept, we constructed a trust belief that models a pervasive adversary and ran
some experiments to examine how trust might improve security in Tor. In particular, we considered
how trust might be used to prevent the first--last correlation attack when accessing a given online
chat service. These experiments just show the potential for improvement from using trust; they
do not take into account other attacks or how to maintain good performance.

We suppose that users are trying to avoid a powerful adversary called ``The Man.'' This
adversary might compromise relay \emph{families} and
AS or IXP \emph{organizations}, where a family or organization is a group controlled by the same
entity. Each family is compromised by the adversary independently with probability between 0.001 and 0.1, where the probability increases as the family's longevity in Tor decreases. Each AS and IXP organization
is compromised independently with probability 0.1.

Against The Man, we examine both how users can choose more-secure paths through Tor and how the
service can choose server locations to make them more securely accessible via Tor.
The algorithm we consider for trust-aware path selection begins by choosing as its \emph{guards}
(\ie, relays used by a client to start all connections into Tor) the three guard relays with the
smallest
probabilities that the adversary observes the path from the client to the guard or the
guard itself. Then for a given destination, the algorithm chooses one of these guards and an
\emph{exit} (\ie, a relay that will initiate connections outside the Tor network) to
minimize the probability of a first--last correlation attack. The algorithm for choosing
server location considers only those ASes containing an exit, which minimizes the chance
for the adversary to observe traffic between the exit and destination. The algorithm
greedily locates each server for the greatest reduction in the probability that
users in the
most common locations (and using the given trust-aware path-selection algorithm)
are open to a first--last correlation attack. Probabilities are estimated by repeated sampling.

For our experiments, we used Web chat server
\texttt{webirc.oftc.net} as the destination service. This IRC service
is run by the Open and Free
Technology Community and is popular with Tor developers. We considered users coming from 58 of the
top ASes as measured by Juen~\cite{juen-thesis}, which in their observations included the client
location for over 95\% of Tor client packets.

Our results are shown in Table~\ref{table:experiments}. The first row shows a first--last compromise
probability of over 0.1 for a client using Tor's default path-selection algorithms to
connect to the current chat server location. We can see that by using trust to choose
guard and exit relays, clients can reduce the compromise probability by a factor of
over 2.8 on average. When in addition the service changes the location of its server, that
probability drops again by a factor of over 2.7 and approaches the minimum possible of
$(0.1)^2=0.01$.
It appears that adding additional server
locations does not add significantly to user security. Note that each probability is estimated
with 100,000 samples, which can explain why some probabilities are slightly below 0.01 and why the probabilities sometimes increase slightly when a server is added.
See App.~\ref{ap:exp} for further experiment details.

\vspace{-\baselineskip}

\begin{table}
\caption{First--last correlation probabilities against The Man for 58 client locations}
\label{table:experiments}
\centering
\begin{tabular}{|l|c|c|c|c|}
\hline
& Mean & Median & Min & Max\\
\hline
Tor default path selection & 0.132 & 0.127 & 0.108 & 0.164\\
\hline
Clients use trust & 0.046 & 0.049 & 0.026 & 0.091 \\
\hline
Clients \& service use trust, 1 server & 0.017 & 0.018 & 0.009 & 0.033\\
\hline
Clients \& service use trust, 2 servers &  0.017 & 0.017 & 0.009 & 0.034\\
\hline
Clients \& service use trust, 3 servers & 0.017 & 0.017 & 0.009 & 0.033\\
\hline
\end{tabular}
\end{table}

\vspace{-2\baselineskip}

\section{Ongoing and Future Work}

Ongoing and future work includes the further development and investigation of Tor path-selection 
algorithms that use trust as formalized here, the further development and analysis of methods to 
express trust that are natural and usable, and the continued analysis of possible trust errors and 
their effects.  Two particularly important tasks are the development of collections of 
trust beliefs that capture important use cases and the study of how users can
use different trust beliefs without being identified by that behavior.

\appendix

\section{Full System Overview}\label{ap:overview}

The system comes with an ontology that describes types of network elements (\eg, AS, link, and relay-operator types), the relationships between them that capture the effects of compromise by an adversary, and attributes of these things.  While we provide an ontology, this may be replaced by another ontology as other types of threats are identified.  Appendix~\ref{ap:ontology} describes the requirements for replacement ontologies.  Roughly speaking, the ontology identifies the types of entities for which the system can automatically handle user beliefs when constructing the Bayesian Belief Network (BBN) for the user.  A user may express beliefs about other types of entities, but she would need to provide additional information about how those entities relate to entities whose types are in the ontology.  The ontology is provided to the user in order to facilitate this.

In general, we expect that the system will provide information about network relationships, such as which ASes and IXPs are on a certain virtual link or which Tor relays are in a given relay family.  We generally expect the user to provide information about human--network relationships such as which individual runs a particular relay.  Note that this means the user might need to provide this type of information in order to make some of her beliefs usable; if she has a belief about the trustworthiness of a relay operator, she would need to tell the system which relays that operator runs in order for the trustworthiness belief to be incorporated into the BBN.

Using the ontology and various published information about the network, the system creates a preliminary ``world'' populated by real-world instances of the ontology types (\eg, specific ASes and network links).  The world also includes relationship instances that reflect which particular type instances are related in ways suggested by the ontology.  User-provided information may include revisions to this system-generated world, including the addition of types not included in the provided ontology and instances of both ontology-provided and user-added types.  The user may also enrich the information about the effects of compromise (adding, \eg, budget constraints or some correlations).  

The user expresses beliefs about the potential for compromise of various network entities; these beliefs may refer to specific network entities or to entities that satisfy some condition, even if the user may not be able to effectively determine which entities satisfy the condition.  This user-provided information is used, together with the edited world, to create a Bayesian Belief Network (BBN) that captures the implications of the user's trust beliefs.  A user may express a belief that refers to an entity or class of entities whose type is in the given ontology.  For such beliefs, the system will be able to automatically incorporate those beliefs into the BBN that the system constructs.  A user may also express beliefs about entities whose types are not included in the ontology.  If she does so, she would need to provide the system with information about how those entities should be put into the BBN that the system constructs.

The system and the user need to agree on the language(s) in which she will express her beliefs.  Different users (or, more likely, different organizations that want to provide collections of beliefs) may find different languages most natural for expressing beliefs.  The language specification(s) must describe not only the syntax for the user but also (\emph{i}) how her structural beliefs will be used in modifying the system-generated world and (\emph{ii}) how her other beliefs will be used to translate the edited world into a BBN.  The user's beliefs may include boolean predicates evaluated on elements in the world.  We sketch a default language for these predicates, but this could be replaced by any other language on which the user and system agree.  

The BBN may be sampled to obtain probabilities of compromise for relays and (virtual) links in the Tor network.

\subsection{Construction sequence}\label{app:conseq}

An overview of the system's actions is below.  The procedure to produce the BBN is treated as a black box.  In reality, it involves many steps, but these depend on the belief language used.  The procedure for the belief language described in App.~\ref{app:slang} is presented in App.~\ref{app:bbntrans}.
\begin{enumerate}
	\item World generation from ontology: $_R W^I_T$
	\begin{itemize}
		\item As described in App.~\ref{app:sysworld}, the system generates a preliminary view of the world based on the ontology and its data sources.  We denote the result by $_R W^I_T$.
		\item This should include system attributes
	\end{itemize}
	\item Augmenting the types with the user's types: $_R W^I_{T'}$
	\begin{itemize}
		\item The user may provide additional types (as a prelude to adding instances of those types to the world).  We use $_R W^I_{T'}$ to denote the augmentation of $_R W^I_T$ by adding the user's types.
	\end{itemize}
	\item Adding user-specified instances of types (ontology and user-provided): $_R W^{I'}_{T'}$
	\begin{itemize}
		\item The user may add instances of any of the types in $_R W^I_{T'}$.  We use $_R W^{I'}_{T'}$ to denote the augmentation of $_R W^I_{T'}$ by adding these new instances and removing any that the user wishes to omit.
	\end{itemize}
	\item Adding user-specified relationships (between instances in $_R W^{I'}_{T'}$): $_{R'} W^{I'}_{T'}$
	\begin{itemize}
		\item The user may specify additional parent/child relationships beyond those included in $_R W^{I'}_{T'}$.  In particular, any new instances that she added in the previous step will not be related to any other instances in the world unless she explicitly adds such relationships in this step.  We use $_{R'} W^{I'}_{T'}$ to denote the augmentation of $_R W^{I'}_{T'}$ by adding these new relationships and by removing any that the user wishes to omit.
	\end{itemize}
	\item Edit system-provided attributes (not budgets or compromise effectiveness)
	\item Add new user-provided attributes
	\item Add budgets
	\item Add compromise effectiveness (this will default to something, perhaps specified in the ontology, if values aren't given; for relationships of types not given in the ontology, we will use a default value unless the user specifies something when providing the relationship instance)
	\item Produce BBN
	\begin{itemize}
		\item The translation process is described in Sec.~\ref{app:bbntrans}.
	\end{itemize}
\end{enumerate}

\section{Full Ontology}\label{ap:ontology}

Before presenting the ontology that we use in this work, we describe our general requirements for ontologies in this framework.  This allows our ontology to be replaced with an updated version satisfying these requirements.  For example, mutual legal assistance treaties (MLATs) are a topic of current interest.  There is presently no suitable source of information about MLATs for our system to use~\cite{cortes14pc}.  If a database of these is developed that can be reliably used to determine automatically the effects of MLATs on the power of state-level adversaries, it would be natural to update the ontology to reflect the system's ability to do this.

\subsection{General requirements for ontologies}

We assume that the any ontology used in our system has the following properties:
\begin{itemize}
	\item It has a collection $\types$ of \emph{types}.  We use the ontology to describe relationships between the types in the ontology.
	\item A collection $\edges$ of (directed) \emph{edges} between types (with $\edges\cap\types = \emptyset$).  The edges are used to specify relationships; if there is an edge from $T_1$ to $T_2$ in the ontology, then the compromise of a network element of type $T_1$ has the potential to affect the compromise of a network element of type $T_2$.
	\item Viewed as a directed graph, $(\types,\edges)$ is a DAG.
	\item A distinguished set of \types\ called the \emph{output types}.  This is for convenience; these are the types of instances that we expect will be sampled for further use.  We generally expect the output types to be exactly the types in the ontology that have no outgoing edges.
	\item Each element of $\types\cup\edges$ has a \emph{label} that is either ``system'' or ``user.''  For an edge $e$ from type $T_1$ to type $T_2$, if either $T_1$ or $T_2$ has the label ``user,'' then $e$ must also have the label ``user.''  These labels will be used to indicate the default source of instances of each type.  (However, the user may always override system-provided information.)
	
Types or edges with the label ``user'' might be natural to include in an ontology when the type/edge is something about which the system cannot reliably obtain information but the ontology designer is able to account for instances of the edge/type in the BBN-construction procedure.
	\item A collection $\attr$ of \emph{attributes}.  Each attribute includes a name, a data type, a source (either ``system'' or ``user'').  Each element of $\types\cup\edges$ may be assigned multiple boolean combinations of attributes; each combination is labeled with either ``required'' or ``optional.''\footnote{In the rest of this document, we assume that each combination is just a single ``optional'' attribute without any connectives.  The semantics of individual attributes depend on the translation procedure that produces the BBN.  We expect that a boolean combination of attributes will be interpreted as possible combinations of attributes that the translation procedure can handle; for example, it might be able to process either a pair of integers or a single real value.  Richer applications of the ``optional'' and ``required'' labels might be allowed as well, although we do not need them here.}
\end{itemize}

\subsection{Our ontology}

Figure~\ref{fig:ontology} depicts the ontology used in our system.  The two ovals at the bottom are the output types: Tor relays and Tor (virtual) links, which include the links between clients and guards and between exits and destinations.  The rounded rectangles correspond to types in the ontology; instances of these will be factor variables in the BBN.  Attributes are depicted as cylinders; the interpretation of these will be described below.   
Filled-in types and solid edges indicate elements and attributes whose label is ``system.''  Unfilled types/attributes and dotted edges indicate elements whose label is ``user.''  As noted above, all attributes in the ontology we present here have the label ``optional.''

\subsubsection{User-provided types}

The types and relationships that are provided by the system in constructing the preliminary world are described in App.~\ref{app:sysworld}.  We describe the others here; instances of these are added by the user in ways specified below.

\begin{description}
	\item[Hosting Service (and incident edges)] Hosting services that might be used to host Tor relays.  If a service hosts a particular relay, there would be a relationship instance from the service to the relay.  If a service is known to be under control of a particular legal jurisdiction or company, the appropriate incoming relationship instance can be added.
	\item[Corporation (and incident edges)] Corporate control of various network elements may be known.  A corporation that is known may be added as an instance of this type.  If the corporation is known to be subject to a particular legal jurisdiction, then a relationship edge from that jurisdiction to the corporation can be added.  Similarly, hosting services, ASes, and IXPs that a corporation controls may be so indicate via the appropriate relationship instances.
	\item[Router/switch/\emph{etc}.] This corresponds to a physical router or switch.  We do not attempt to identify these automatically, but ones known to the user (or a source to which the user has access) may be added as instances of this type.
	\item[Physical connection] Particular physical connections, such as a specific cable or wireless link, may be known and of interest.  
	\item[(Physical connection, Virtual link)] If a virtual link is known to use a specific physical connection, then that can be reflected in a relationship between the two.
\end{description}

\subsubsection{Attributes}

The attributes in our ontology are depicted by cylinders in Fig.~\ref{fig:ontology}.  The two at the box in the top right can be applied to all non-output type instances, so we do not explicitly show all of the types to which they can be applied.

\begin{description}
	\item[System-generated attributes] These include relay-software type and physical location.  Users may edit these, \eg, to provide additional information.
	\item[Connection type] This is an attribute of physical-connection instances.  It is represented as a string that describes the type of connection (\eg, \texttt{"submarine cable"}, \texttt{"buried cable"}, or \texttt{"wireless connection"}).  A user would express beliefs about connection types; if the type of a connection is covered by the user's beliefs, then the probability of compromise would be affected in a way determined by the belief in question.
	\item[Budget] This attribute, which is supplied by the user at her option, may be applied to any non-output type instance.  There are two variants.  Both are represented as an integer $k$ and another value.  In the first variant, the other value is a type; in the second variant, the other value is the string \texttt{"all"}.  Multiple instances of this attribute may be applied to a single type instance as long as they have distinct second values; if one of these is the second variant, then all others will be ignored.  This allows the user to express the belief that, if the type instance is compromised, then its resources allow it to compromise $k$ of its children.  In the first variant of this attribute, the instance may compromise $k$ of its children of the specified type (and perhaps $k'$ of its children of a different type, if so specified).  In the second variant of this attribute, the instance may compromise $k$ of its children across \emph{all} types.\footnote{The resources needed to compromise instances of different types may vary widely.  However, we include the second variant so that a budget that covers all of an instance's children can be modeled in some fashion.}
	
	As discussed below, we must approximate the effects of resource constraints so that the BBN can be efficiently sampled.
	\item[Region] This is an attribute of legal jurisdiction.  It is represented as a boolean predicate on geographic coordinates. 
	\item[Compromise effectiveness] This attribute is syntactically similar to the budget attribute.  It is supplied by the user at her option for instances of any non-output type, and there are effectively two variants.  This is represented as a probability $p\in[0,1]$ and a boolean predicate on type instances; we distinguish non-trivial predicates from the always-true predicate $\top$.  Multiple instances of this attribute may be applied to a single type instance as long as no two non-$\top$ predicates evaluate to True on the same input.  Only one instance of this attribute with $\top$ may be present; if it is, then all other instances of the attribute for the type instance are ignored.
	
	This attribute allows the user to express beliefs about the effect of compromise of one type instance on its children, either uniformly or according to type.  For example, a compromised AS might attempt to compromise all of its routers, but make a mistake in in the configuration file that it copies to each router of a certain hardware model.  This might happen with probability $p = 10^{-4}$.  Absent budget restrictions, either all or none of the routers of this model will be compromised \emph{due to this action by the AS}.  This is in contrast to the effects of budgets.
	\item[Router/Switch Kind] This is an attribute of routers/switches and is represented as a set of strings.  We expect the user to use this to describe aspects of routers/switches that she might know about and want to use in her trust beliefs, \eg, the model number or firmware version of specific routers and switches.
	\item[Relay Hardware] This is an attribute of relays and is represented in the same way as the router/switch kind.  Also analogously to that attribute, we expect that the user would use this to describe aspects of relay hardware that she might know about and potentially use in her trust beliefs.
\end{description}

\subsection{System-generated world}\label{app:sysworld}
The system provides users with a \emph{world} consisting of
\emph{type instances} and \emph{relationship instances}
that are consistent with the types and relationships specified in the ontology.
Formally, a world is a DAG in which each vertex
is a type instance, each edge is a relationship instance, and an attribute function
assigns each vertex a vector of attributes.
A type instance represents a real-world object of the
specified type. For example, ``AS3356'' is a type instance of the AS type, and
``Level 3 Communications'' is a type instance of the AS Organization type. A
relationship instance will only relate two instances of types that are related
in the ontology. For example, (Level 3 Communications, AS3356) is an instance
of the (AS Organization, AS) relationship type and indicates that AS3356 is a
member of Level 3 Communications. The attributes of a type instance provide information
that users can incorporate into their trust beliefs, such as the location of a given Tor
relay. The world can be modified by users in ways provided by the trust language.  We assume that each instance has a unique identifier and an indication of the type of which it is an instance.

The system generates a world as follows:
\begin{enumerate}
\item The current Tor consensus and the server descriptors it references are
used to create the following instances and attributes, which concern relays:
    \begin{itemize}
    \item \textbf{Tor Relay}: An instance is created for each relay in the
    consensus.
    \item \textbf{Relay Family}: An instance is created for connected component
    of relays, where two relays are connected if they mutually reference each
    other in their descriptors~\cite{tor-pathspec}.
    \item \textbf{(Relay Family, Tor Relay)}: An instance of this relationship
    is created for each relay belonging to a given family.
    \item \textbf{Relay Software Type}: This attribute is added to each relay based on the
    operating system reported in the relay's descriptor.
    \end{itemize}
\item 
Standard techniques~\cite{usersrouted-ccs13} are used to construct an AS-level Internet routing map.  Data that can be used to create such a map
includes the CAIDA internet topology~\cite{caida-topology}, the CAIDA AS
relationships~\cite{caida-asrels}, and RouteViews~\cite{routeviews}. This map
is then used to create the following instances:
    \begin{itemize}
    \item \textbf{Virtual Link}: An instance is created representing the
    path between each Autonomous System and \emph{entry guard} as
    well as between each Autonomous System and \emph{exit relay}.
    An entry guard is a Tor relay that satisfies the requirements to
    serve as the relay that a Tor client directly connects to. An exit relay
    is any relay that can be used as the relay that connects directly to 
    a client destination. Entry guards and exit relays are determined from
    the Tor consensus. A virtual link instance
    represents both directed paths between the Autonomous System and relay, which
    may differ due to Internet route asymmetries~\cite{asymmetry-globecom05}.
    \item \textbf{AS}: An instance is created for each AS observed in the
    RouteViews data.
    \item \textbf{(AS, Virtual Link)}: An instance of this relationship is created for
    each AS that appears on the path in either direction between the virtual link's AS and
    its relay, as determined by the Internet routing map.
    \end{itemize}
\item Internet Exchange Points (IXPs) are added to paths in the AS-level
Internet map based on data from the IXP Mapping
Project~\cite{ixp-mapping-imc09}. These additions are are used to create the
following instances:
    \begin{itemize}
    \item \textbf{IXP}: An instance is created for each IXP that appears on at
    least one path in the Internet map.
    \item \textbf{(IXP, Virtual Link)}: An instance of this relationship is created for
    each IXP that appears on the path in either direction between the virtual link's AS
    and its relay, as determined by the Internet routing map.
    \end{itemize}
\item ASes are clustered into organizations using the results of Cai~\etal.~\cite{Cai12b}, and IXPs are clustered into organizations using the results
of Johnson~\etal.~\cite{usersrouted-ccs13}. Each cluster represents a single
legal entity that controls multiple ASes or IXPs, such as a company. The
clusters are used to create the following instances:
    \begin{itemize}
    \item \textbf{AS Organization}: An instance is created for each AS cluster.
    \item \textbf{IXP Organization}: An instance is created for each IXP
    cluster.
    \item \textbf{(AS Organization, AS)}: An instance of this relationship is created for
    each AS in a given AS cluster.
    \item \textbf{(IXP Organization, IXP)}: An instance of this relationship is created
    for each IXP in a given IXP cluster.
    \end{itemize}
\item The system provides physical locations and legal jurisdictions for several of the 
ontology types. IP location information, such as from the MaxMind GeoIP
database~\cite{maxmind-geoip},
provides location information for entities with IP addresses. The location of
IXPs is frequently available on the Web as well~\cite{ixp-mapping-imc09}. These
data are used to create the following instances and attributes:
    \begin{itemize}
    \item \textbf{Legal jurisdiction}: An instance of this type is created for each
    country.
    \item \textbf{(Legal jurisdiction, Relay)}: An instance of this relationship is
    created for each relay in a given country, as determined by the relay's IP address
    and the IP location information.
    \item \textbf{(Legal jurisdiction, IXP)}: An instance of this relationship is
    created for each IXP in a given country, as determined by the IP addresses of the
    IXP or other public IXP information.
    \item \textbf{Physical location}: This attribute is added to each relay with its
    geographic coordinates (\ie, latitude and longitude), as determined from its IP
    address. This attribute is also added to each IXP with its geographic coordinates,
    based on its IP addresses or other public IXP information.
    \end{itemize}
\end{enumerate}

\section{User Beliefs}

\subsection{Types of Beliefs}

Broadly, users may have two types of beliefs: those about the structure of the network, \etc., and those about trust.  Beliefs of the first type are used by the system to edit the preliminary, system-generated world to produce an ``edited world.''  (We expect that this should be done once, not on a per-adversary basis.)  Beliefs of the second type are given as input, together with the edited world, to the procedure that produces the BBN.  We call these two types of beliefs ``structural beliefs'' and ``trust beliefs,'' respectively.  
We now turn to the description of structural and trust beliefs.  We then describe an example language that a user may use to express her beliefs, and then we describe how to translate beliefs represented in this language into a BBN.

\subsection{Structural beliefs}

A user may have beliefs about instances of types and edges from the ontology.  For types, a user may believe that an instance of that type exists; her belief about that instance must include a unique identifier for the instance and any required attributes.  
This type instance is then added to the system-generated world.  The type of the instance may be system-generated, in which case this belief represents an edit to the system-generated world, or it may be user-generated.  If the instance's type is user-generated, then the user must describe to the system how the instance should be translated to the BBN that the system produces from the edited world.  
	
	For edges, a user may believe that one type instance is the parent of another type instance.  Her belief about such a relationship must include any required attributes of the corresponding edge type in the ontology.  This relationship instance is then added to the system-generated world.  If the edge type is not part of the ontology, the user must describe how the edge affects the computation of values in the BBN that the system produces.

\subsection{Trust beliefs}

A user may also have beliefs about the probability of compromise of any factor.  The BBN construction that we describe below supports two general types of trust beliefs: absolute beliefs, which state an absolute probability of compromise, and relative beliefs, which modify the computation of compromise probability.

If the system allows a user to provide trust beliefs about a class of network elements or relationships, then she may also provide default beliefs that override the system's defaults for that class.

\subsubsection{A language for predicates}\label{app:predlang}

We expect that the user may want to express some of her beliefs (trust and perhaps also structural) in terms of predicates, even though she might not be able to effectively evaluate these herself.  For example, the user's trust in ASes with very few routers might be different than her trust in ASes with many routers (perhaps because she believes that larger ASes are more likely to have processes, policies, and organizational experience that prevent misconfiguration).  She might capture this with a predicate that expresses whether the number of routers in an AS (in the edited world) is at least as great as a specified threshold.

The belief languages must thus incorporate a language for predicates that the system can interpret.  We treat the predicate language as a separate component, and we sketch here one predicate language that will be used by all of our example belief languages.  This language includes:
\begin{description}
	\item[Connectives and operators] Basic logical connectives (including negation)
	\item[Typing] Testing whether an instance or attribute is or is not of a specified type; users may test for types not in the ontology (\eg, to check types that they have added)
	\item[Sets] Sets (explicitly enumerated or defined by some predicate) and set mem\-ber\-ship/non-membership
	\item[Membership] A predicate may depend on a set and test whether a value is in that set.
	\item[Tests of attribute values] Tests must be appropriate to the data type used in the attribute; equality and inequality tests are allowed unless specified otherwise.  Predicates may test user-defined attributes.\footnote{We expect that user-defined attributes will only be tested by the user, \eg, through predicates that she specifies on those attributes.  As noted in the construction sequence in App.~\ref{app:conseq}, the system will not change the structure of the world based on user-defined attributes.}  This may reference user-defined attributes.  
	\item[Tests of the world structure (in trust beliefs only)] After the world is constructed and edited (\ie, when applying trust beliefs but not when applying structural beliefs), we allow predicates in beliefs to refer to the structure of the world.  
\end{description}

\newcommand{\augtypes}{\ensuremath{\mathcal{T}'}}
\newcommand{\insts}{\ensuremath{\mathcal{I}}}
\newcommand{\auginst}{\ensuremath{\mathcal{I}'}}
\newcommand{\rels}{\ensuremath{\mathcal{R}}}
\newcommand{\augrels}{\ensuremath{\mathcal{R}'}}

\subsection{Sample language}\label{app:slang}

We now describe a sample language for users' structural and trust beliefs.  This incorporates predicates, which might be expressed using the predicate language just outlined.  In general, we assume that there is a set \vals\ of values that the user may use to express levels of trust.  We illustrate this here by taking \vals\ to be $\{\scb,\lcb,\unk,\ltb,\stb\}$; we think of these as ``Surely Compromised,'' ``Likely Compromised,'' ``Unknown,'' ``Likely Trustworthy,'' and ``Surely Trustworthy.''  Our examples will not rely on \vals\ having exactly five elements, but we think this is one natural way that users might think about their trust in network elements.

\subsubsection{Structural beliefs}

Let $\rels$ be the set of relationship instances in the system-created world.  \augrels\ will be \rels\ augmented with all of the user-specified relationships.  

\begin{description}
	\item[Novel types]  A user may define new types via expressions of the form $(\mathtt{''ut''},\linebreak[4] tname, struct_{req}, struct_{opt})$, where $\mathtt{''ut''}$ is a string literal, $tname$ is a string (the name of the type) that must be distinct from all other $tname$ values the user specifies and from all elements of $\types$, and where $struct_{req}$ and $struct{opt}$ are both descriptions of data structures (these may be empty data structures, which might be indicated by $\mathtt{NULL}$).

We write $\augtypes$ for the set containing the elements of $\types$ together with all of the $tname$ values provided by the user.

	\item[Type instances] An ordered list of tuples $(T,D,n,P,C)$, where $T\in\augtypes$, $D$ is a data structure that is valid for $T$, and $n$ is a unique identifier among these tuples.\footnote{We assume that the system provides unique identifiers for the system-generated type instances and that the values of $n$ in the user's list of tuples are distinct from those identifiers.}  
	
We write \auginst\ for the set formed by augmenting \insts\ with these new instances.
	\item[Relationship instances] A set of pairs $(p,c)$, where $p$ and $c$ are type instances from \auginst.\footnote{We abuse notation and use $p$ and $c$ in place of the unique identifiers associated with each type instance in the edited world.}  We do not need to specify new relationship types, only the additional relationship instances.
\end{description}

\subsubsection{Trust beliefs}

\begin{description}
	\item[Relative beliefs] These are beliefs of the form $(s,\mathsf{P},v)$, where $s$ is a string other than $\mathtt{''abs''}$, $\mathsf{P}$ is a predicate on factor variables, and $v\in\vals$.
	
	Note that, in our translation procedure below, relative beliefs affect the probability of compromise of a factor in the BBN that is not otherwise compromised through the causal relationships captured in the world.
	\item[Absolute beliefs] These are beliefs of the form $(\mathtt{''abs''},\mathsf{P},v)$, where $\mathsf{P}$ is a predicate on factor variables and $v\in\vals$.  A belief such as this says that the chance a variable satisfying $\mathsf{P}$ is compromised is captured by $v$.  Note that it is the user's responsibility to ensure that no two different absolute beliefs have predicates that are simultaneously satisfied by a node if those beliefs have different values for $v$.  We do not specify what value is used if this assumption is violated.\footnote{A natural approach is to allow the use to specify these in an ordered list and using the last satisfied predicate.}
	\item[Budget] Expressed as either $(\mathtt{''bu1''},I,T,k)$ or $(\mathtt{''bu2''},I,\mathtt{''all''},k)$, where $\mathtt{''bu1''}$ and $\mathtt{''bu1''}$ are string literals, $I$ is a type instance in the edited world, $T$ is a type in the edited world, and $k$ is an integer.  The interpretation is that, in expectation, compromise of the node with this attribute will lead to compromise of $k$ of its children (of type $T$ in the first variant, or of all its children in the second variant).
	\item[Compromise effectiveness] Expressed as either $(\mathtt{''ce1''},I,\mathsf{P}_\mathsf{ce},v)$ or $(\mathtt{''ce2''},I,\linebreak[4]\top,v)$, where $\mathtt{''ce1''}$ and $\mathtt{''ce2''}$ are string literals, $I$ is an instance of a non-output type in the edited world, $\mathsf{P}_\mathsf{ce}$ is a on instances of a fixed type, $\top$ is a distinguished symbol, and $v\in\vals$.  The interpretation is that, if instance $I$ is compromised, then it compromises its children satisfying $\mathsf{P}_\mathsf{ce}$ (or all children, if $\top$ is given) with probability corresponding to $v$.  

The likely range of compromise-effectiveness probabilities may differ from the likely range of other compromise probabilities.  In a language such as we are describing here, the values $\stb$, \etc., might have different corresponding probabilities for compromise-effectiveness than they do for absolute and relative beliefs.  Another alternative is to allow users to specify a probability $p\in[0,1]$ instead of a value $v\in\vals$ as the last element of these tuples.
\end{description}

\subsection{Translations to BBNs}\label{app:bbntrans}

A translation procedure in general needs to take the edited world (reflecting the structural beliefs and attribute values provided by the user) and the user's trust beliefs as input and produce a BBN as output.  The output variables of the BBN should match the nodes in the edited world that are instances of types designated as output types in the ontology or the user's structural beliefs.  Here, we present a translation procedure that fits with the rest of the system we describe (it matches our particular ontology, \etc.).

\subsubsection{Our translation procedure}

Let $W'$ be the final world that appears in the construction sequence described above.

\begin{itemize}
	\item For each node (type instance) in $W'$, the BBN contains a corresponding variable.  We refer to the BBN variable by the same name as the node in $W'$.
	\item For each compromise-effectiveness belief $B = (s,n,\mathsf{P},v)$ about a node $n$, there is a corresponding child $v_B$ of $n$ in the BBN.  The table for $v_B$ is such that, if $n$ is uncompromised, then $v_B$ is uncompromised; if $n$ is compromised, then $v_B$ is compromised with probability $p(v)$ and uncompromised otherwise.  (We use $p(v)$ to denote the probability value that the system assigns to the value $v\in\vals$ that is part of the user's belief language.)\ \ The children of $v_B$ in the BBN are the nodes in the BBN that correspond to nodes in $W'$ that (1) are children of $n$ and (2) satisfy the predicate $\mathsf{P}$ from the belief $B$.  Assign these edges the weight set $\{1\}$.
	
	If there are children of $n$ in $W'$ that do not satisfy any of the predicates in the compromise-effectiveness beliefs about $n$ (including, \eg, when the user has no compromise-effectiveness beliefs), then make these nodes children of $n$ in the BBN.  Assign to each of these edges the singleton weight set whose element is the appropriate default probability.\footnote{We assume that there are default values---perhaps just a single, common one---for the probability that the compromise of a node leads to the compromise of its children.  These values might naturally depend on the types involved.  Here, we suggest $1$ as a common default value.}
	
	\item For each budget belief $B = (s,n,\mathsf{P},k)$ about a node $n$, let $c_{n,\mathsf{P}}$ be the number of children of $n$ (in $W'$) that satisfy $\mathsf{P}$.  For each of these children, in the BBN, replace the single value in the edge's weight set by that value multiplied by $k/c_{n,\mathsf{P}}$.
	
	\item Assign to each non-CE-belief node $n$ a ``risk set'' $R_n$ that is initially empty.  For each belief $B = (s,\mathsf{P},v)$ that has not already been evaluated and whose initial entry is not $\mathtt{''abs''}$, if $n$ satisfies $\mathsf{P}$, then add $v$ to $R_n$ (retaining duplicates, so that $R_n$ is a multiset).
	
	\item Construct the tables for each non-CE node in the BBN.  (We have already constructed the tables for the CE-belief nodes.)  Let $n$ be a non-CE node.  For each subset $\mathcal{S}$ of $n$'s parents, if $S$ is the multiset of weights on the edges from nodes in $\mathcal{S}$ to $n$, and if $R$ is the multiset of risk weights associated with $n$, then the probability that $n$ is compromised given that its set of compromised parents is exactly $\mathcal{S}$ is:
\[
1 - \left(\prod_{p\in S}(1-p)\right)\left(\prod_{q\in R}(1-q)\right).
\]
Note that, if the user has no parents, then the first product will be empty (taking a value of $1$), and the probability of compromise will be determined solely by the risk factors unless the user expresses beliefs that override these.

	\item If the user provides a belief $B = (\mathtt{''abs''},\mathsf{P},v)$, then nodes satisfying $\mathsf{P}$ are disconnected from their parents.  Their compromise tables are then set so that they are compromised with probability $p(v)$ and uncompromised with probability $1-p(v)$.  This allows a user to express absolute beliefs about factor variables in the BBN (hence ``$\mathtt{abs}$'').  In particular, she may express beliefs about input variables whose compromise would otherwise be determined by their attributes.  
\end{itemize}

\subsubsection{Potential extensions}

We assume that adversaries are acting independently, although this may not always be the case.  One natural example of inter-adversary dependence occurs with the compromise of resource-constrained instances in the world.  For example, an ISP's resources may limit it to monitoring $k$ of its routers.  If both the ISP and the country (or other legal jurisdiction) controlling it are a user's adversaries, then they should compromise the same set of the ISP's routers.  (This is true whether we model this compromise probabilistically, with $k$ routers compromised in expectation, or through some other means.)  This might be modeled statically by changing the structure of the BBN, but dynamic compromise and more general inter-adversary dependence may require other approaches.

At this point, our system does not include instances in the world in constructs that correspond to cities or states/provinces.  These are most naturally viewed as instances of legal jurisdictions, and the user may well have beliefs about the corresponding laws or enforcement regimes.  One way that we envision the user may address these is by adding to the world instances of legal jurisdictions that carry a ``Boundary'' attribute, effectively a predicate that can be evaluated on the system-provided geolocation data.  The system could then determine which network entities are in which of these user-supplied jurisdictions. Physical locations might be handled this way as well, as long as the location is ``large enough'' relative to the resolution of the geolocation process.

Mutual legal assistance treaties (MLATs) concern the exchange of information between countries about possible violations of the laws of a participating country.  If a user has a state-level adversary, then an MLAT between the adversary country and another country might effectively compromise the second country.    
As noted above, it may be natural to add these to the ontology once suitable related sources of information become available.  The BBNs that we presently construct could be extended to include MLATs by adding two additional layers of variables.  One would contain a variable for each MLAT known to the system; the children of these variables would be the country variables (in the presently constructed BBN) corresponding to countries that are obligated by the respective MLATs to act as adversaries.  The other added layer would contain a new variable for each country; the children of any one of these variables would be all of the MLATs that obligate other countries to provide information to the parent country.  The inherent compromise of countries would be reflected in the top layer; this would propagate through the MLAT layer to effectively compromise other countries, and the rest of the BBN would behave as it does presently.

\subsection{Five-valued example}

The following examples of beliefs illustrate how a user might express her beliefs in our five-valued example language.  We suggest that the compromise probabilities corresponding to the values $\scb$, $\lcb$, $\unk$, $\ltb$, and $\stb$ might be taken to be $0.999$, $0.85$, $0.5$, $0.15$, and $0.02$, respectively.

\begin{enumerate}
	\item Countries in set $S_1$ are likely trustworthy 
	\item Countries in set $S_2$ are likely compromised 
	\item Countries in set $S_3$ are surely compromised
	\item AMS-IX points are likely trustworthy
	\item MSK-IX points are of unknown trustworthiness
	\item Relay family $F_1$ is likely compromised 
	\item Relay family $F_2$ is surely uncompromised
	\item Relay operator $O_1$ is surely uncompromised
	\item Relay operator $O_2$ is likely uncompromised
	\item Hosting company $H_1$ is surely trustworthy 
	\item Submarine cables are of unknown level of trustworthiness
	\item Wireless connections are likely compromised
	\item Relays running Windows are of uncertain trustworthiness (system gets this from relay descriptors)
	\item If an AS is compromised, then it is expected to be able to compromise $4$ of the links that it is on
\end{enumerate}

\section{BBNs}

Bayesian Belief Networks have both strengths and weaknesses as a component of our system.  Their general strengths of being concise, being efficiently sampleable, and allowing computation of other properties of the distribution (\eg, marginal probabilities and maximum likelihood values) are beneficial in our system.  BBNs are especially well-suited to our approach here because of the close structural similarity between our revised worlds and the BBNs we construct from these.

As a disadvantage, BBNs do not represent hard resource constraints efficiently; we can only approximate those here by constraining resources in expectation.  More generally, other negative correlations may be difficult at best to capture, but it is possible that users will hold beliefs that imply negative correlations between compromise probabilities.  

The purpose of this system is to produce an efficiently sampleable representation of compromise probabilities.  Other representations of distributions could also be used, but they might be most naturally generated from trust beliefs in different ways.  A detailed discussion of such approaches is beyond the scope of this work.

\section{Obtaining Trust Beliefs}
We propose that a collection of default beliefs be distributed with this system.  As noted in Sect.~\ref{sec:trust}, this collection would be designed to provide adequate protection for general users.  Users with particular concerns might use other collections of beliefs; these could be provided by, \eg, government entities, privacy organizations, political groups, journalism-focused organizations, or organizations defending abuse victims.

\section{Experimental Results}\label{ap:exp}
\subsection{Constructing The Man}
To construct The Man adversary, we must create a routing map of the Internet that includes ASes,
IXPs, and Tor relays. We must also group ASes and IXPs into organizations, identify relay families,
and
evaluate the longevity of Tor relays. We do so using the techniques and data sources described in
Appendix~\ref{app:sysworld}.

To build the routing map, we used CAIDA topology and link
data from 12/14 and RouteViews data from 12/1/14. The resulting map included
46,368 ASes, 279,841 links between ASes, and 240,442 relationship labels. To group ASes by the 
organization that controls them, we used the results of Cai~\etal.~\cite{Cai12b}. These included
data about 33,824 of the ASes in our map, and they resulted in 3,064 organizations that included
more than one AS with a maximum size of 81 and a median size of 2. We used the results of
Augustin~\etal.~\cite{ixp-mapping-imc09} to identify IXPs and their locations between pairs of ASes.
These results show 359 IXPs and 43,337 AS-pairs between which at least one IXP exists. We then used
the results of Johnson~\etal.~\cite{usersrouted-ccs13} to group IXPs into organizations. These
produce 19 IXP organizations with more than one IXP, for which the maximum size is 26 and
the median size is 2.

We add relays to the routing map using Tor consensuses and descriptors from Tor
Metrics~\cite{tormetrics}. We used the Tor consensus of 12/1/14 at 00:00. The network at this
time included 1,235 relays that were guards only, 670 relays that were exits only, and 493 relays
that were both guards and exits. The consensus grouped relays into 152
families of size greater than one, of which the maximum size was 25 and the median size was 2.
Family uptime was computed as the number of assignments of the \textsf{Running} flag to family
members, averaged over the family members and the consensuses of 12/2014. We mapped
the Tor guards and exits to ASes using Routeviews prefix tables from 12/1/14, 12/2/14, and
11/30/14, applied in that order, which was sufficient to obtain an AS number for all guards and
exits. Note that we observed one exit relay that mapped to an AS that didn't appear in our map,
and so we added that additional AS. There were 699 unique ASes among the guards and exits.

We created paths from each AS in our map to each guard and exit AS. The median number of paths
that we could infer to a guard or exit AS was 46,052 (out of the 46,369 possible).
The maximum AS path
length was 12, and the median AS path length was 4. The maximum number of IXPs on a path was 18, and
the median number was 0.

The resulting BBN for The Man thus included 2398 relay variables (one for each guard and exit) and
32,411,931 virtual links (one from each AS to each guard or exit AS). For any path missing from
our routing map, we simply took the path to include only the source AS and destination AS. The
probability of compromise for a family $f$ with uptime $u_f$ was set to be
$(0.1 - (0.1-0.001)) u_f$.

\subsection{Experiment algorithms}
For all of our experiments, we considered security from 58 of the 60 most common client ASes as measured by Juen~\cite{juen-thesis} (AS8404 and AS20542 did not appear in our map). Juen reports
that these 58 ASes covered 0.951 of client packets observed. In addition, for all of our
experiments,
the compromise probability (\ie, the probability of a first--last correlation attack by The Man) was 
estimated by sampling from The Man BBN (and from Tor's relay selection distribution in
the default case) 100,000 times and using the fraction of compromised samples as 
the probability.

The experiments were conducted as follows:
\begin{itemize}
\item \textbf{Tor default path selection}: For each of our 58 client locations, we choose an exit
and guard using Tor's path-selection algorithm as implemented in
TorPS~\cite{usersrouted-ccs13}. Note that (among other considerations) this does ensure that the
guard and exit don't share
the same family or /16 subnet. Then we sample The Man BBN to determine if the resulting circuit to
the server is vulnerable to a first--last correlation attack.
\item \textbf{Clients use trust}: Guards are chosen for each client location to be the three relays
with the smallest probabilities that the adversary compromises the guard or an AS or IXP on the
path to the guard. To compute the compromise probability of a connection from a given client
location to a given 
destination, we consider using each of the client location's three guards with each Tor exit relay, 
estimate the compromise probability, and choose the lowest resulting probability.
\item \textbf{Service uses trust}: We consider each AS containing an exit relay as a possible 
location for the server. For each server location, we compute the probability of compromise for each 
client location. This is estimated for a given client location by considering each of its guards,
considering each exit sharing the server location, estimating the compromise probability, and using
the minimum of these probabilities. We choose the server location with the minimum average 
compromise over all client locations. For each additional server, we repeat the same process except 
that we only update the compromise probability for a client location if it decreases when using the 
new potential server location.
\end{itemize}

\end{document}